\documentclass[conference]{IEEEtran}
\IEEEoverridecommandlockouts
\usepackage{cite}
\usepackage{amsmath,amssymb,amsfonts}
\usepackage{algorithmic}
\usepackage{graphicx}
\usepackage{textcomp}
\usepackage{xcolor}
\usepackage{multirow}
\usepackage{caption} 
\usepackage{tabularx}

\usepackage[pagebackref,breaklinks,colorlinks]{hyperref}
\hypersetup{colorlinks,linkcolor={blue},citecolor={blue},urlcolor={blue}}  

\def\BibTeX{{\rm B\kern-.05em{\sc i\kern-.025em b}\kern-.08em
    T\kern-.1667em\lower.7ex\hbox{E}\kern-.125emX}}
    
\begin{document}

%
%

\title{Single and Multi-Speaker Cloned Voice Detection: From Perceptual to Learned Features \thanks{This work was partially funded by a grant from the UC Berkeley Center For Long-Term Cybersecurity (CLTC), an award for open-source innovation from the Digital Public Goods Alliance and United Nations Development Program, and from an unrestricted gift from Meta. The public codebase can be found at https://github.com/audio-df-ucb/ClonedVoiceDetection.}
}

\author{\IEEEauthorblockN{Sarah Barrington${^1}$, Romit Barua${^1}$, Gautham Koorma${^1}$, Hany Farid$^{1,2}$}
\IEEEauthorblockA{\textit{School of Information$^{1}$, Electrical Engineering and Computer Sciences$^{2}$},
\textit{University of California, Berkeley}\\
Berkeley, CA USA \\
\{sbarrington, romit\_barua, gautham.koorma, hfarid\}@berkeley.edu}
}

\maketitle

\begin{abstract}
Synthetic-voice cloning technologies have seen significant advances in recent years, giving rise to a range of potential harms. From small- and large-scale financial fraud to disinformation campaigns, the need for reliable methods to differentiate real and synthesized voices is imperative. We describe three techniques for differentiating a real from a cloned voice designed to impersonate a specific person. These three approaches differ in their feature extraction stage with low-dimensional perceptual features offering high interpretability but lower accuracy, to generic spectral features, and end-to-end learned features offering less interpretability but higher accuracy. We show the efficacy of these approaches when trained on a single speaker's voice and when trained on multiple voices. The learned features consistently yield an equal error rate between $0\%$ and $4\%$, and are reasonably robust to adversarial laundering.
\end{abstract}


\section{Introduction}

Computational techniques for modifying a recorded voice to sound like another person while preserving the original semantic meaning--voice conversion--predates today's deepfakes and generative AI by some $65$ years~\cite{sisman2020overview}. The semiannual voice conversion challenge\footnote{\url{http://vc-challenge.org}} evaluates voice cloning submissions on naturalness (rated from 1 = completely unnatural to 5 = completely natural) and speaker identity (rated on a scale of ``same, absolutely sure,'' ``same, not sure,'' ``different, not sure,'' or ``different, absolutely sure''). In the first challenge of 2016, the best-performing system received an average of $3.0$ on the five-point naturalness scale and $70\%$ of the samples were judged on identity to be ``same.'' In 2018, the best-performing system received an average $4.1$ naturalness score, and $80\%$ of the samples were judged on identity to be ``same.'' In 2020, the best naturalness scores continued to hover around $4.0$, but identity ratings were nearly perfect.

Over the past few years, AI-powered voice synthesis has continued to improve (in terms of naturalness and identity), culminating this year in dramatic breakthroughs. Perhaps most striking is zero-shot, multi-speaker text-to-speech (ZS-TTS)\footnote{\url{https://edresson.github.io/YourTTS}} for cloning a voice identity not seen during training from a few seconds to minutes of reference audio~\cite{casanova2022yourtts}. Also striking is the easy access to these voice-cloning technologies through low-cost commercial services\footnote{\url{http://https://beta.elevenlabs.io}}.


While these advances are a major success of the research community, they have also come at a price. Reports of phone scams have emerged in which a call purportedly from a family member claims they were in an accident, arrested, or kidnapped after which the scammer takes over in an attempt to extort money~\cite{npr2023panickycall,cnn2023aiscam}. Similar reports have emerged that financial institutions using voice identification can now be spoofed with voice cloning~\cite{vice2023howibroke}. And, fake audio is adding to already existing problems of disinformation~\cite{ap2023fakeaudio}.

From these disinformation campaigns to small- and large-scale fraud and to the continued erosion in trust of all digital media, it is critical that we develop techniques to distinguish the real from the fake. 

Detection strategies fall into two general categories: (1) active techniques which, at the point of synthesis, embed a perceptible or imperceptible watermark into~\cite{9689555}, or extract a perceptual fingerprint~\cite{9689555} from, synthetically-generated content. These watermarks/fingerprints can then be used to identify content once it is released into the wild; and (2) in the absence of watermarks/fingerprints, passive techniques detect a range of statistical to semantic inconsistencies in synthetically-generated content (see Section~\ref{sec:related-work}).

Our efforts fall into the second category where we describe three related passive approaches for distinguishing real from cloned voices using handcrafted perceptual, generic spectral, or learned features. The benefit of the perceptual features is that they afford a low-dimensional, explainable classifier, while the learned features generally afford better classification performance, with the spectral features affording a compromise between these. These different approaches (Section~\ref{sec:methods}) are evaluated (Section~\ref{sec:results}) against two different real audio datasets and three cloned audio datasets (Section~\ref{sec:datasets}). 

We consider two basic scenarios in which the three feature sets are trained to distinguish real from cloned voices of a single speaker (Section~\ref{sec:results-single-speaker}) and trained simultaneously from multiple speakers (Section~\ref{sec:results-multi-speaker}).

\subsection{Related Work}
\label{sec:related-work}

By way of background, Almutairi and Elgibreen~\cite{almutairi2022review} provide a review and a quantitative comparison of various audio deepfake detection methods, and the First International Workshop on Deepfake Detection for Audio Multimedia focused on synthetic audio detection~\cite{yi2022add}. In this section, we highlight a few of these approaches and those most closely related to ours.

Classical approaches for detecting synthetic speech typically exploit statistical differences between synthetic and human speech. Ogihara et al.~\cite{ogihara2005discrimination}, for example, proposed a technique that exploits differences in pitch between synthetic and human speech. De Leon et al.~\cite{de2012synthetic} extended this work by exploiting additional pitch features including stability and jitter. In addition to these pitch differences, they also observed that the transition between phonemes occurs more rapidly in synthetic speech. AlBadawy et al.~\cite{albadawy2019detecting} showed that synthetic speech contains specific and unusual higher-order spectral correlations that are not typically found in human speech. 

Moving beyond these statistical approaches, more recent approaches have incorporated explicit vocal and perceptual models. Blue et al.~\cite{blue2022you} employed fluid-dynamic models to estimate the arrangement of the vocal tract during speech generation, and argued that synthetic speech yields unlikely anatomical structures. Li et al.~\cite{li2022comparative} compared $16$ physical and perceptual features for synthetic audio detection and highlighted the importance of perceptual features. They found that in noisier conditions where the quality of the synthetic audio is low, the perceptual linear prediction technique~\cite{hermansky1990perceptual}, which combines spectral analysis with linear prediction analysis, outperforms other features. They also analyzed the distribution of these features for real and synthetic speech, providing useful benchmarks for selecting discriminative features.

Variations in prosody have also been used to detect synthetic audio. For example, Attorresi et al.~\cite{attorresi2022combining} combined a speaker embedding representing distinct voice features (e.g.,~timbre and pitch contour) with a prosodic embedding representing variational style (e.g.,~rhythm and intonation). Their experiments on the ASVspoof19 dataset show that a combination of these two embeddings yields a $3-15$ percentage point improvement in equal error rate (EER) over baseline models (RawNet2, MFCC-ResNet, Spec-ResNet).

End-to-end deep learning has also been deployed to identify synthetically-generated speech. Muller et al.~\cite{muller2022does}, for example, evaluated the generalizability of various deepfake detection algorithms of $12$ end-to-end architectures, and tested them on a novel in-the-wild (IWA) dataset of public figures collected from social networks and video-streaming platforms\footnote{\url{https://deepfake-demo.aisec.fraunhofer.de/in_the_wild}}. They observed that the raw audio-based end-to-end models outperformed the feature-based models, with the RawNet2 model proposed by Tak et al.~\cite{tak2021end} achieving the lowest equal error rate (EER) of $3.2\%$ on the ASVspoof19 dataset and an EER of $33.9\%$ on the IWA dataset (with chance performance at $50\%$).

Lastly, Pianese et al.~\cite{pianese2022deepfake} evaluated the use of various off-the-shelf speaker verification tools for synthetic voice detection and found them effective and robust to intentional and unintentional laundering (e.g.,~transcoding, resampling, etc.). This approach yielded an average EER of $15.0\%$ on the ASVspoof19, FakeAVCeleb, and IWA datasets.

Most forensic approaches seek to distinguish real from synthetic voices regardless of identity. A more personalized biometric approach can also be taken in which a person's distinct voice characteristics are used to distinguish the real from the fake~\cite{pianese2022deepfake}.

Beyond classifying speech as synthetic or real, recent efforts have also focused on identifying fingerprints that can identify specific synthesis architectures~\cite{yan2022initial}. And, although somewhat outside of the scope of our work, there has also been an effort to detect audio spoofing in the form of a rebroadcast attack in which a person's voice is recorded and replayed~\cite{tom2018end,tak2021end}. 

We take a hybrid approach in terms of the audio features--leveraging learned, spectral, and perceptual features--and in terms of considering both single-speaker (personalized) detectors and multi-speaker (non-personalized) detectors. We evaluate our detectors on a number of real and cloned voices and evaluate the vulnerability to standard laundering attacks.

\begin{table}[t]
\centering
\resizebox{0.45\textwidth}{!}
{
\begin{tabular}{|l|l|c|c|}
\multicolumn{4}{c}{SINGLE-SPEAKER} \\
\hline
{\bf Type}      & {\bf Name}        & {\bf Clips (\#)} & {\bf Duration (sec)} \\
\hline
\hline
Real      & LJSpeech    & $13{\small,}100$  & $86{\small,}117$ \\
\hline
Synthetic & WaveFake    & $91{\small,}700$  & $603{\small,}081$ \\
          & ElevenLabs  & $13{\small,}077$  & $78{\small,}441$ \\
          & Uberduck    & $13{\small,}094$  & $83{\small,}322$ \\
\hline
\multicolumn{4}{c}{~}\\
\multicolumn{4}{c}{MULTI-SPEAKER} \\
\hline
{\bf Type}      & {\bf Name}       & {\bf Clips (\#)} & {\bf Duration (sec)} \\
\hline
\hline
Real      & TIMIT       & $4{\small,}620$    & $14{\small,}192$ \\
\hline
Synthetic & ElevenLabs    & $5{\small,}499$   & $15{\small,}413$ \\
\hline
\end{tabular}
}
\caption{An overview of the real and synthetic datasets used in our single-speaker (top) and multi-speaker (bottom) evaluations. The $91{\small,}700$ WaveFake samples correspond to $13{\small,}100$ samples per each of seven different vocoder architectures, hence the larger number of clips and duration.}
\label{tab:datasets}
\vspace{-0.25cm}
\end{table}
\begin{figure*}[t!]
\centering
    \begin{tabular}{l}
    \hspace{1cm} REAL \\
      \includegraphics[width=0.93\textwidth]{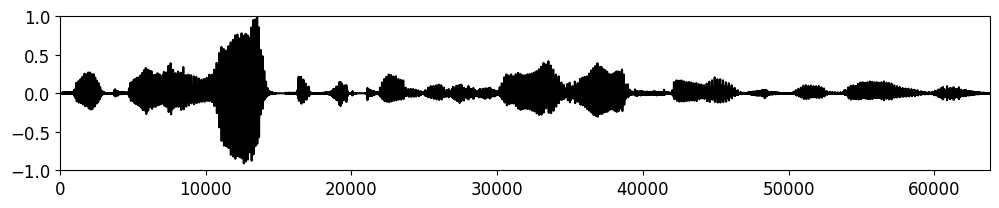} \\
    \hspace{1cm} SYNTHETIC \\
      \includegraphics[width=0.93\textwidth]{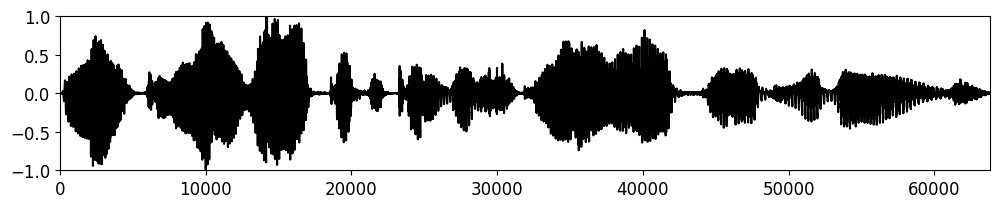} 
    \end{tabular}
  \vspace{-0.25cm}
  \caption{Example real audio (top) and synthetic audio (bottom) temporal waveforms (each normalized into the amplitude range $[-1,1]$) for the same utterance. Note the difference in the length of the silences and the differences in overall amplitude and amplitude modulation over time.}
  \label{fig:audio-samples}
  \vspace{-0.25cm}
\end{figure*}
%
%

\section{Datasets}
\label{sec:datasets}

A selection of publicly available datasets was used to develop and test our models (see Table~\ref{tab:datasets}). For the evaluation of single-speaker detection, the LJSpeech~\cite{ljspeech17} and WaveFake datasets~\cite{frank2021wavefake} were used. The LJSpeech dataset is a publicly available dataset consisting of $13{\small,}100$ short audio clips of a single female speaker, Linda Johnson, reading passages from seven non-fiction books. The WaveFake dataset\footnote{\url{https://github.com/RUB-SysSec/WaveFake}} comprises $117{\small,}985$ audio clips generated from the LJSpeech dataset using seven different vocoder architectures. Linda Johnson's voice was cloned from the LJSpeech dataset using the leading commercial text-to-speech (TTS) platforms ElevenLabs and Uberduck. Each transcript from the LJSpeech corpus was re-generated in the cloned voice. 

 
For the evaluation of our perceptual features (Section~\ref{sec:methods}) and multi-speaker detection (Section~\ref{sec:results-multi-speaker}), we used the TIMIT dataset\cite{timit}, consisting of $462$ real male and female American-English speakers, uttering a total of $1{\small,}718$ different phonetically-rich sentences \cite{timit}. Each of these phrases was fed to ElevenLabs with one of $11$ distinct voices: nine of the voices were built into ElevenLabs, and we cloned the remaining two voices to mimic Presidents Biden and Obama using $1{\small ,}038$ and $1{\small ,}192$ seconds of audio recordings. The resulting dataset provided a diverse range of real and synthesized voices with a one-to-one correspondence of the underlying utterances. To ensure balanced representation, utterances with only one human speaker were removed from the dataset, and the remaining audio clips were randomly sampled to select clips with the greater count of the real or synthetic voice per utterance. This process yielded a total of $763$ real and $763$ synthesized audio clips. Lastly, each real and synthesized audio was normalized into the amplitude range $[-1,1]$.

All audio files were downsampled to 16khz and the seven WaveFake architectures were randomly sampled such that the total number of WaveFake clips were equal to that of Uberduck and ElevenLabs. 

\section{Methods}
\label{sec:methods}

We describe three approaches for classifying speech as synthetic or real (single class), and for identifying the underlying synthesis architecture (multi class). These approaches range from low-dimensional (and interpretable) handcrafted features to higher-dimensional generic spectral audio features, to even higher-dimensional (and less interpretable) learned neural features. The next three sections describe these features followed by a description of a simple classifier that ingests these features for the purpose of single- and multi-class classification.

\subsection{Perceptual}

Shown in Fig.~\ref{fig:audio-samples} is a pair of real (top) and synthetic (bottom) waveforms (each normalized into the amplitude range $[-1,1]$) for the same utterance (\textit{``nuclear rockets can destroy airfields with ease''}) from which we can see some qualitative differences. For the same utterance, the real human voice shows a lower average normalized amplitude and higher amplitude variability. And, we observe that real voices exhibit more frequent and noticeable pauses between certain words. Using the real and fake TIMIT dataset, and as described next, we designed a set of handcrafted features to determine if these simple temporal-domain observations would yield reliable classification between real and synthetic audio.

{\em Pause:} A pause in the temporal waveform is identified as a segment of audio with $100$ consecutive samples with a rolling average amplitude less than $0.5\%$ of the maximum normalized amplitude (all audios are normalized into the range $[-1,1]$). 

The mean/standard deviation of pause length (as a percentage of the audio length) for real and synthetic audio contained within the TIMIT dataset is $27.27/8.49$ and $13.57/6.56$. A two-sided t-test reveals a strong statistical difference in these distributions ($p \ll 10^{-10}$).

We quantify these differences by extracting four summary statistics from the identified pauses: the pause ratio (the ratio of pauses relative to the length of the audio), the mean pause length (specified as the number of samples), the standard deviation of pause length, and the number of pauses (the number of pauses, of course, depends on the number of words per utterance, but our training dataset consisted of the same utterances for both real and synthetic audio).

{\em Amplitude:} Two amplitude features are extracted capturing the consistency and variation in voices. To begin, the absolute values of each waveform are temporally smoothed with a fifth-order Butterworth low-pass filter. From this smoothed waveform, we compute the overall mean amplitude and mean amplitude of the temporal derivative. The mean/standard deviation of mean amplitude for real and synthetic audio contained within the TIMIT dataset is $0.06/0.02$ and $0.10/0.02$ ($p \ll 10^{-10}$), again showing a significant difference.


\subsection{Spectral}

For generic spectral features, we employed the openSMILE library (speech \& music interpretation by large-space extraction)~\cite{eyben2010opensmile}. For an arbitrary-length audio clip, openSMILE generates $6{\small,}373$ scalar-valued features such as summary statistics (mean, standard deviation, etc.), regression coefficients, linear predictive coding coefficients, and various peak-related functionals. A simple dimensionality reduction (SelectFromModel\footnote{\url{https://scikit-learn.org/stable/modules/generated/sklearn.feature\_selection.SelectFromModel.html}}) was used to reduce the number of features to $20$.

\subsection{Learned}

For the end-to-end learned audio features, we employed Nvidia's open-source TitaNet model~\cite{koluguri2022titanet}. TitaNet  was initially trained for speaker identification using end-to-end additive margin angular loss, which enhances the separation of speaker identity in the latent space. Using an encoder-decoder architecture, TitaNet converts 16KHz sampled raw audio files into $192$-D embeddings. We treat these embeddings as features for the downstream classification task. 


\subsection{Classification}

For each of the three feature sets described above, we employed a linear logistic regression and a non-linear random forest classifier for a single-class (real vs. synthetic) or multi-class (real vs. specific synthesis architecture) task. In each case, the full data set was split into a $60\%$ training, $20\%$ validation (for hyper-parameter tuning), and $20\%$ testing. All results below are for the testing portion of the dataset.

\section{Results}
\label{sec:results}

We describe classification accuracy for a personalized, single-speaker task in which a classifier is trained on learned, spectral, or perceptual features for a single-speaker identity. We next describe the generalization of these classifiers to a multi-speaker task in which a classifier is trained across multiple speakers. The classifiers are evaluated against the generated voices, and laundered voices. Lastly, we compare our results to a ElevenLabs' detector.

\begin{table*}[t!]
\centering
\resizebox{0.99\textwidth}{!}{
\begin{tabular}{ |c|c|c|c|c|c|c|c|c|c|c|}
 \multicolumn{11}{c}{SINGLE-SPEAKER} \\
 \hline
 \textbf{Dataset} & \textbf{Model} & \multicolumn{3}{|c|}{\textbf{Synthetic Accuracy (\%)}} & \multicolumn{3}{|c|}{\textbf{Real Accuracy (\%)}} & \multicolumn{3}{|c|}{\textbf{EER (\%)}}\\
  \hline
   &  &  \textbf{Learned} & \textbf{Spectral} & \textbf{Perceptual} &  \textbf{Learned} & \textbf{Spectral} & \textbf{Perceptual} &  \textbf{Learned} & \textbf{Spectral} & \textbf{Perceptual}\\
 \hline
 \hline
 EL       & single (L)  & 100.0 & 99.2 & 78.2 & 100.0 & 99.9 & 72.5 & 0.0 & 0.5 & 24.9 \\
          & single (NL) & 100.0 & 99.9 & 82.2 & 100.0 & 100.0 & 80.4 & 0.0 & 0.1 & 18.6 \\
\hline          
 UD       & single (L) & 99.8 & 98.9 & 51.9 & 99.9 & 98.9 & 54.0 & 0.1 & 1.1 & 47.2 \\
          & single (NL) & 99.7 & 99.2 & 54.4 & 99.9 & 99.0 & 56.5 & 0.2 & 0.9 & 44.5 \\
 \hline
  WF      & single (L) & 96.5 & 78.4 & 57.8 & 97.1 & 82.3 & 45.6 & 3.3 & 19.7 & 48.5 \\
          & single (NL) & 94.5 & 87.6 & 50.3 & 96.7 & 90.2 & 52.7 & 4.4 & 11.2 & 48.6 \\
\hline
 EL+UD    & single (L) & 99.7 & 94.8 & 63.4 & 99.9 & 97.1 & 60.3 & 0.2 & 4.2 & 37.9 \\
          & single (NL) & 99.7 & 99.2 & 57.3 & 99.9 & 99.6 & 69.0 & 0.2 & 0.8 & 37.6 \\
 \hline
 EL+UD+WF & single (L) & 93.2 & 79.7 & 58.4 & 98.7 & 93.0 & 57.6 & 3.6 & 15.9 & 42.1 \\
          & single (NL) & 91.2 & 90.6 & 53.1 & 99.0 & 94.1 & 64.7 & 4.1 & 7.9 & 41.6 \\
 \hline
 \hline
 EL+UD    & multi (L) & 99.9 & 96.6 & 61.0 & 100.0 & 94.6 & 35.7 & - & - & - \\
          & multi (NL) & 99.7 & 98.3 & 65.6 & 100.0 & 97.2 & 43.2 & - & - & - \\
 \hline
 EL+UD+WF & multi (L) & 98.8 & 80.2 & 45.1 & 97.3 & 64.3 & 22.9 & - & - & - \\
          & multi (NL) & 98.1 & 94.2 & 48.6 & 96.3 & 84.4 & 27.6 & - & - & - \\
\hline
\multicolumn{11}{c}{~}\\
\multicolumn{11}{c}{SINGLE-SPEAKER: ADVERSARIAL LAUNDERING} \\
 \hline
 \textbf{Dataset} & \textbf{Model} & \multicolumn{3}{|c|}{\textbf{Synthetic Accuracy (\%)}} & \multicolumn{3}{|c|}{\textbf{Real Accuracy (\%)}} & \multicolumn{3}{|c|}{\textbf{EER (\%)}}\\
  \hline
   &  &  \textbf{Learned} & \textbf{Spectral} & \textbf{Perceptual} &  \textbf{Learned} & \textbf{Spectral} & \textbf{Perceptual} &  \textbf{Learned} & \textbf{Spectral} & \textbf{Perceptual}\\
 \hline
 \hline
 EL       & single (L)  & 95.5 & 94.3 & 61.1 & 94.5 & 92.6 & 65.2 & 4.9 & 6.7 & 36.6 \\
          & single (NL)  & 96.0 & 96.2 & 70.4 & 95.4 & 95.6 & 69.6 & 4.1 & 4.1 & 30.1 \\
\hline          
 UD       & single (L) & 95.4 & 81.1 & 61.4 & 91.8 & 84.3 & 44.7 & 6.3 & 17.3 & 46.7 \\
          & single (NL) & 95.4 & 86.8 & 52.9 & 93.3 & 86.1 & 55.9 & 5.5 & 13.6 & 45.6 \\
 \hline
  WF      & single (L) & 87.6 & 60.7 & 59.6 & 85.0 & 70.4 & 42.5 & 13.9 & 34.4 & 49.4\\
          & single (NL) & 83.6 & 77.1 & 51.4 & 85.6 & 76.7 & 53.9 & 15.3 & 23.1 & 47.3\\
\hline
 EL+UD    & single (L) & 95.2 & 79.1 & 54.0 & 91.7 & 78.4 & 59.8 & 6.2 & 21.3 & 43.1 \\
          & single (NL) & 94.8 & 86.1 & 55.2 & 93.3 & 90.0 & 62.4 & 6.0 & 12.0 & 41.4 \\
 \hline
 EL+UD+WF & single (L) & 83.7 & 70.9 & 50.6 & 88.6 & 72.9 & 59.7 & 13.2 & 28.2 & 44.8\\
          & single (NL) & 83.4 & 79.2 & 53.0 & 90.7 & 85.1 & 60.7 & 12.5 & 17.9 & 43.6 \\
 \hline
 \hline
 EL+UD    & multi (L) & 94.2 & 85.6 & 50.9 & 91.0 & 77.1 & 29.1 & - & - & - \\
          & multi (NL) & 94.5 & 91.7 & 53.2 & 90.3 & 82.9 & 41.3 & - & - & - \\
 \hline
 EL+UD+WF & multi (L) & 89.8 & 65.4 & 35.3 & 83.1 & 44.3 & 26.2 & - & - & - \\
          & multi (NL) & 88.8 & 78.8 & 39.8 & 82.1 & 63.0 & 28.6 & - & - & - \\
\hline
\multicolumn{11}{c}{~}\\
\multicolumn{11}{c}{MULTI-SPEAKER} \\
\hline
 \textbf{Dataset} & \textbf{Model} & \multicolumn{3}{|c|}{\textbf{Synthetic Accuracy (\%)}} & \multicolumn{3}{|c|}{\textbf{Real Accuracy (\%)}} & \multicolumn{3}{|c|}{\textbf{EER (\%)}}\\
  \hline
   &  &  \textbf{Learned} & \textbf{Spectral} & \textbf{Perceptual} &  \textbf{Learned} & \textbf{Spectral} & \textbf{Perceptual} &  \textbf{Learned} & \textbf{Spectral} & \textbf{Perceptual}\\
 \hline
 \hline
 EL       & single (L)  & 100.0 & 94.2 & 83.9 & 99.9 & 98.3 & 86.9 & 0.0 & 3.0 & 13.1 \\
          & single (NL)  & 92.3 & 96.3 & 82.2 & 100.0 & 99.7 & 87.7 & 0.1 & 1.6 & 13.7 \\
\hline
\end{tabular}
}
\caption{Accuracy for a personalized, single-speaker classification of unlaundered audio (top) and audio subject to adversarial laundering in the form of additive noise and transcoding (middle). Shown in the bottom table is the non-personalized, multi-speaker accuracy. Dataset corresponds to ElevenLabs (EL), Uberduck (UD), and WaveFake (WF); Model corresponds to a linear (L) or non-linear (NL) classifier, and for a single-classifier (real v. synthetic) or multi-classifier (real vs. specific synthethis architecture; accuracy (\%) is reported for synthetic audio, real audio, and (for the single-classifiers) equal error rate (EER).}
\vspace{-0.25cm}
\label{tab:results}
\end{table*}
%
%

\subsection{Single Speaker}
\label{sec:results-single-speaker}

Shown in Table~\ref{tab:results} (top panel) is the accuracy for distinguishing real from synthetic audio (model: single) and real from specific synthetic audio architecture (model: multi) using a linear (model: L) and non-linear (model: NL) classifier, evaluated against single or multiple datasets (ElevenLabs [EL], Uberduck [UD], WaveForm [WF]). Each column corresponds to the accuracy for correctly classifying real and synthetic audio using the learned, spectral, or perceptual features. The far-right columns report the equal error rate (the EER is the point on the receiver operating curve (ROC) where the false acceptance rate (incorrectly classifying a synthetic voice as real) and false rejection rate (incorrectly classifying a real voice as synthetic) are equal).

As expected, the non-linear classifier generally affords better accuracy. For the spectral features, for example, across all dataset combinations the non-linear classifiers afford an average $4.1$ percentage point reduction in EER.

Accuracy on the learned features outperforms the spectral and perceptual features, with an average EER on single datasets (and linear classifier) of $0.0\%$, $0.1\%$, and $3.3\%$ for the learned features as compared to $0.5\%$, $1.1\%$, and $19.7\%$ for the spectral features, and $24.9\%$, $47.2\%$, and $48.5\%$ for the perceptual features. 

Generally speaking, classifiers trained and tested on a single dataset (EL, UD, or WF) perform better than those trained on two or more datasets. And, accuracy on the single-class task is higher than on multi-class.

\subsection{Laundering}
\label{subsec:laundering}

To test the robustness of our methods against unintentional or intentional adversarial laundering attacks, we split our real and synthetic datasets into four equal classes consisting of the unlaundered audio, the unlaundered audio corrupted with additive Gaussian noise with an SNR sampled uniformly between $10$ and $80$dB, the unlaundered audio transcoded (AAC) at a bitrate of 64K, 127K, or 196K, and the unlaundered audio transcoded and corrupted with noise. Shown in Table~\ref{tab:results} (middle panel) are the resulting classification accuracies in the same format as described above. 

As expected, laundering degrades classification accuracy. The spectral features were particularly impacted which is perhaps not surprising since the additive noise and transcoding introduce broad-band spectral distortions. 

As compared to the unlaundered voices, the EER for the learned features jumps by $7.5$ percentage points for the linear classifier and $6.9$ percentage points for the non-linear classifier.

\subsection{Multi Speaker}
\label{sec:results-multi-speaker}

The above results are based on personalized classifiers trained to distinguish real from synthetic audio for a specific individual. 
Shown in the lower panel of Table~\ref{tab:results} is the accuracy for a multi-speaker classifier trained and tested on the TIMIT-ElevenLabs dataset. This classifier is trained to detect synthetic voices regardless of the underlying identity. The learned features yield similar EER as compared to single speaker and the spectral EER is only slightly higher. The perceptual features, on the other hand, yield a lower EER dropping from $18.6\%$ to $13.7\%$  (for the nonlinear classifier). We hypothesize that this improvement is because the cadence for the single speaker (LJ) as she is reading is highly structured, as compared to a more conversational style. Regardless, these results imply that our features are not speaker specific, but seem to capture synthesis artifacts regardless of identity.

\subsection{Comparison}
\label{sec:comparison}

ElevenLabs recently released a classifier designed to determine if an audio sample was generated by their synthesis engine\footnote{\url{https://beta.elevenlabs.io/blog/ai-speech-classifier}}. With a reported accuracy of ${\small >}99\%$ accuracy for unlaundered samples and ${\small >}90\%$ accuracy for laundered samples, this classifier is on par with our classifier based on learned features (Table~\ref{tab:results}, top and middle panels, row EL). We tested the ElevenLabs classifier on a random sample of $50$ real and $50$ ElevenLabs synthesized audio samples, each laundered with additive Gaussian noise and transcoded at varying compression levels (see Section~\ref{subsec:laundering}). Classification accuracy was perfect, as compared to our average accuracy of $95.8\%$ using the learned features and non-linear classifier. Despite this slightly lower performance, our classifier, unlike the ElevenLabs classifier, can detect samples from other synthesis engines: we verified that ElevenLabs mis-classifies synthetically-generated audio from Uberduck and WaveFake.

Although comparison to other published techniques is difficult due to differences in the underlying training and tresting datasets, generally speaking we achieve lower or equal EERs to the techniques described in Section~\ref{sec:related-work}.

\section{Discussion}

In the field of digital forensics, image- and video-based techniques have outpaced those of audio forensics. And for good reason. Until fairly recently synthetic voices were not particularly natural or identity-preserving. This, however, is no longer the case and it is now possible to create highly natural and realistic voices from only a few minutes of a person's voice. When coupled with increasingly high-quality deepfake videos, it is quickly becoming possible to create highly realistic deepfake videos of anyone saying anything.

Combining video and audio analyses (e.g.,~\cite{bohavcek2022protecting}) offers the advantage of a richer data source and more chances to detect statistical or semantic inconsistencies. Purely audio-based techniques, however, are needed to contend with phone-based scams and fake leaked audios of world or corporate leaders.

While low-dimensional, interpretable features are attractive, it is clear that the end-to-end learned features afford better discrimination. We did not combine all three features because the learned features significantly outperformed the others.

The advantage of a single-speaker approach is that it can learn highly specific and distinct speaking styles that are difficult for a synthetic voice to perfectly mimic. The drawback is that, unlike multi-speaker techniques, it does not scale well to protect a large number of possible victims of voice cloning. We see the need for both single- and multi-speaker approaches. Our results suggest that the same underlying feature selection and classification can be adapted for both tasks.

As new voice synthesis architectures emerge, it will be important for forensic techniques to generalize across new architectures. Our results suggest that this type of generalization is possible, but that performance generally degrades as the classifier is tasked with categorizing voices from increasingly more diverse synthesis architectures. To the extent that the goal is to distinguish real from synthetic voices, a single-class approach can be taken. It may be informative, however, to also refine multi-class approaches in which the classifier is able to specify which synthesis architecture was used to generate a fake voice; such information could be useful in tracking down the source of disinformation campaigns or illegal activities.

As our field continues to develop techniques for distinguishing real from fake content, we encourage those on the synthesis side to help mitigate potential abuse from deepfakes by embedding imperceptible watermarks into synthetically generated content (see, for example, Adobe's Content Authenticity Initiative\footnote{https://contentauthenticity.org}). While this is not a panacea, it, along with the types of forensic techniques described here, will take us a long way to mitigating abuse from AI-generated content.


\bibliographystyle{ieee_fullname}
\bibliography{main}


\end{document}